\documentclass{aa}
\usepackage{graphicx}
\begin{document}

\title{Characterising the magnetic fields of the Herbig Ae/Be stars HD\,97048, HD\,150193, HD\,176386, and MWC\,480\thanks
{Based on observations obtained at the European Southern Observatory (ESO programme 085.C-0137(A)) and SOFIN observations at the 
2.56-m Nordic Optical Telescope on La Palma.}}

\author{
S. Hubrig\inst{1}
\and
M. Sch\"oller\inst{2}
\and
I.~Ilyin\inst{1}
\and
C.~R.~Cowley\inst{3}
\and
Z.~Mikul\'{a}\v{s}ek\inst{4,5}
\and
B.~Stelzer\inst{6}
\and
M.~A.~Pogodin\inst{7,8}
\and
R.~V.~Yudin\inst{7,8}
%\and
%J.~F.~Gonz\'alez\inst{9}
\and
M.~Cur\'e\inst{9}
}

\institute{
Leibniz-Institut f\"ur Astrophysik Potsdam (AIP), An der Sternwarte 16, 14482 Potsdam, Germany\\
\email{shubrig@aip.de}
\and
European Southern Observatory, Karl-Schwarzschild-Str.\ 2, 85748 Garching bei M\"unchen, Germany
\and
Department of Astronomy, University of Michigan, Ann Arbor, MI 48109-1042, USA
\and
Department of Theoretical Physics and Astrophysics, Masaryk University, Brno, Czech Republic
\and
Observatory and Planetarium of J. Palisa, V{\v S}B - Technical University, Ostrava, Czech Republic
\and
INAF-Osservatorio Astronomico di Palermo, Piazza del Parlamento 1, 90134 Palermo, Italy
\and
Pulkovo Observatory, Saint-Petersburg, 196140, Russia
\and
Isaac Newton Institute of Chile, Saint-Petersburg Branch, Russia
%\and
%Instituto de Ciencias Astronomicas, de la Tierra, y del Espacio (ICATE), 5400 San Juan, Argentina
\and
Departamento de F\'isica y Astronom\'ia, Facultad de Ciencias, Universidad de Valpara\'iso, Chile
}

\abstract
 % context heading (optional)
{Our knowledge of the presence and the role of magnetic fields in intermediate-mass pre-main-sequence stars remains very poor. }  
{
We present the magnetic properties of four Herbig Ae/Be stars that have 
not been previously studied in detail.
}
%New determinations of the mean longitudinal magnetic field for four Herbig Ae/Be stars
%are presented. 
{
%For three 
Our results for the three Herbig Ae/Be stars HD\,97048, HD\,150193, and HD\,176386
are based on multi-epoch low-resolution spectra obtained in spectropolarimetric mode
with FORS\,2 mounted on the VLT.
New high-resolution polarimetric spectra
were obtained for MWC\,480 with the SOFIN spectrograph installed at the Nordic
Optical Telescope. 
We discuss 41 FORS\,2 low-resolution observations of the Herbig Ae/Be stars HD\,97048, HD\,150193, and HD\,176386 and determine their 
rotational periods. Using stellar fundamental parameters 
and the longitudinal magnetic field phase curves, we place constraints on the magnetic field geometry.
Three high-resolution circularly polarised SOFIN spectra obtained for MWC\,480 were measured using the moment technique
where wavelength shifts between right- and left-hand side
circularly polarised spectra are interpreted in terms of a longitudinal 
magnetic field  $\left<B_z\right>$.
}
{
Our search for periodicities resulted in  $P=0.693$\,d for HD\,97048, $P=1.317$\,d for HD\,150193, and  $P=0.899$\,d for
HD\,176386. The magnetic field geometry can likely be described 
by a centred dipole with a polar magnetic field strength $B_{\rm d}$ of several hundred Gauss.
%1.5 and 2\,kG and an inclination $\beta$ of the magnetic axis to the rotation axis of 84$\pm$13$^{\circ}$.
The longitudinal magnetic-field measurements of MWC\,480 reveal the presence of a strong kG field, which was undetected 
in our previous low-resolution polarimetric observations with FORS\,1. 
A weak magnetic field was detected in the circumstellar
components of the \ion{Ca}{ii} H\&K lines and  the \ion{Na}{i}~D lines, indicating a complex interaction between the stellar
magnetic field and the circumstellar environment.
}
{}


\titlerunning{Magnetic fields of Herbig Ae/Be stars}
\authorrunning{S.\ Hubrig et al.}
\maketitle

%________________________________________________________________

\section{Introduction}

Spectropolarimetric observations of several Herbig Ae/Be stars have indicated that
magnetic fields are important ingredients of the intermediate-mass star formation process. 
Models of magnetically driven accretion and outflows
successfully reproduce many observational properties of low-mass
pre-main sequence stars (the classical T\,Tauri stars), but the picture is less clear for higher mass stars.
It is important to understand how the magnetic fields in Herbig Ae/Be stars are  generated and how these fields
interact with the circumstellar environment, displaying a combination of disk, wind, accretion, and jets.
To date, these fields have not been studied for a representative number of Herbig Ae/Be stars.
%In spite of the importance of the studies of magnetic fields for the proper 
%understanding of how the magnetic fields in Herbig Ae/Be stars are generated and how these fields interact 
%with the circumstellar environment, presenting a combination of disk, wind, accretion, and jets, 
%these fields have not been studied for a representative number of Herbig Ae/Be stars.
Magnetically controlled behaviour has been found in both the spectroscopic and photometric variability
of the strongly magnetic Herbig Ae star HD\,101412 (Hubrig et al.\ \cite{Hubrig2010,Hubrig2011}).
The small amount of  UVES spectra acquired for this star have been analysed uncovering
 variations in equivalent widths, radial velocities, 
line widths, line asymmetries, and mean magnetic field modulus over the rotation period of 42.1\,d.
Magnetic field configurations are obviously of utmost importance to understanding
 the magnetospheres of Herbig Ae/Be stars.
%and their interaction with the circumstellar environment presenting a combination of disk, wind, accretion, and jets. 
Since the accreting material has to be lifted from the disk midplane to 
higher stellar latitudes near the stellar photosphere, some coupling
between the accreting plasma and a stellar magnetic field 
is required. Progress in understanding the disk-magnetosphere interaction can, however, only come 
from studying a sufficient number of targets in detail to look for patterns encompassing this type of pre-main sequence stars.

In this work, we present series of mean longitudinal magnetic-field measurements for the Herbig Ae/Be stars HD\,97048, HD\,150193, and 
HD\,176386
obtained at low resolution with the multi-mode instrument FORS\,2 at the VLT, and describe their magnetic field geometries.
Magnetic fields of the order of 120--250\,G were detected in these stars for the first time during our visitor run 
with FORS\,1 in May 2008 (Hubrig et al.\ \cite{Hubrig2009}).
The follow-up program developed to search for magnetic field 
variations over the rotation cycle was allocated for observations with FORS\,2 in 2010. 
The previous spectropolarimetric observations of the star MWC\,480 (=HD\,31648) with FORS\,1 
%was previously observed with FORS\,1 only once, 
revealed the presence of a weak longitudinal magnetic field 
$\left<B_{\rm z}\right> = 87\pm 22$\,G and distinct circular polarisation signatures in spectral lines
originating in the circumstellar (CS)
environment (Hubrig et al.\ \cite{Hubrig2006,Hubrig2007}). We report here on three new  high-resolution polarimetric 
spectra obtained with the SOFIN spectrograph installed at the Nordic Optical Telescope, which confirm the presence of a 
magnetic field in this star in both photospheric and CS lines.
%It is generally accepted that accretion from a disk is an integral
%phase of star formation. 
%A number of Herbig Ae stars and classical T\,Tauri stars are surrounded 
%by active accretion disks and, probably, most of the excess emission seen at
%various wavelength regions can be attributed to the interaction of the disk
%with a magnetically active star (e.g.\ Muzerolle et al.\ \cite{Muzerolle2004}).
%This interaction is generally referred to as
%magnetospheric accretion. Recent magnetospheric accretion models for these stars 
%assume a dipolar magnetic field
%geometry and accreting gas from a circumstellar disk falling ballistically 
%along the field lines onto the stellar surface. 

\section{Magnetic field measurements and period determination}
%\subsection{Spectroscopic material}

\begin{table}
\centering
\caption{
Target stars with multi-epoch spectropolarimetric observations.
}
\label{tab:targetlist}
\centering
\begin{tabular}{llrl}
\hline
\hline
\multicolumn{1}{c}{Object \rule{0pt}{2.6ex}} &
\multicolumn{1}{c}{Other} &
\multicolumn{1}{c}{V} &
\multicolumn{1}{c}{Spectral} \\
\multicolumn{1}{c}{Name} &
\multicolumn{1}{c}{Identifier} &
 &
\multicolumn{1}{c}{Type} \\
\hline
MWC\,480     & HD\,31648     & 7.7 & A3Ve   \\
HD\,97048    & CU\,Cha       & 8.5  & A0pshe    \\
HD\,150193   & V2307\,Oph    & 8.9  & A1Ve     \\
HD\,176386   & CD$-$37 13023 & 7.3  & B9V    \\
%\hline
%\multicolumn{4}{c}{Magnetic Herbig Ae/Be stars from previous studies}\\
%\hline
%HD\,31648  & MWC\,480    & 7.7 & A3pshe     \\
\hline
\end{tabular}
%\begin{flushleft}
%Notes:
\tablefoot{
Spectral types and visual magnitudes are taken from SIMBAD.
}
%\end{flushleft}
\end{table}

\begin{table}
\caption[]{
Magnetic field measurements of Herbig Ae/Be stars with FORS\,1/2.
%All quoted errors are 1$\sigma$ uncertainties.
%The FORS\,1 measurements  
%published by Hubrig et al.\ (\cite{Hubrig2009}) are presented in the first line, respectively.
%Logbook of the spectropolarimetric observations of , and of the
%magnetic field measurements.
%Results of our magnetic field measurements.
}
\label{tab:log_meas}
\begin{center}
\begin{tabular}{ccr@{$\pm$}lr@{$\pm$}l}
\hline \hline\\[-7pt]
%\tableline\tableline
\multicolumn{1}{c}{MJD} &
\multicolumn{1}{c}{Phase} &
\multicolumn{2}{c}{$\left<B_{\rm z}\right>_{\rm all}$ [G]} &
\multicolumn{2}{c}{$\left<B_{\rm z}\right>_{\rm hyd}$ [G]} \\
%\multicolumn{2}{c}{$\left<B_{\rm z}\right>_{\rm all}$} &
%\multicolumn{2}{c}{$\left<B_{\rm z}\right>_{\rm hyd}$} \\
%\multicolumn{1}{c}{} &
%\multicolumn{1}{c}{} &
%\multicolumn{2}{c}{[G]} &
%\multicolumn{2}{c}{[G]} \\
\hline\\[-7pt]
%\tableline
\multicolumn{6}{c}{HD\,97048} \\
%\tableline
\hline\\[-7pt]
  54609.137 & 0.083 &    164 &  42 &    188 & 47 \\
  55279.131 & 0.349 &  $-$92 &  41 & $-$135 & 58 \\
  55311.109 & 0.467 & $-$163 &  53 & $-$195 & 59 \\
  55312.035 & 0.802 &   $-$6 &  74 &     12 & 91 \\
  55320.094 & 0.426 &  $-$44 &  54 &  $-$62 & 62 \\
  55322.119 & 0.346 & $-$123 &  42 &  $-$62 & 48 \\
  55324.105 & 0.210 &      4 &  59 &      9 & 63 \\
  55326.987 & 0.367 & $-$124 &  34 & $-$156 & 44 \\
  55334.039 & 0.537 & $-$154 &  53 & $-$193 & 61 \\
  55335.119 & 0.094 &    127 & 120 &    135 & 49 \\
  55345.987 & 0.768 &     54 &  59 &     26 & 63 \\
  55347.180 & 0.488 & $-$186 &  61 & $-$231 & 64 \\
  55348.070 & 0.772 &  $-$36 &  38 &  $-$12 & 64 \\
  55349.102 & 0.260 &   $-$7 &  74 &     65 & 78 \\
%\tableline
\hline\\[-7pt]
\multicolumn{6}{c}{HD\,150193}\\
%\tableline
\hline\\[-7pt]
  54609.092 & 0.118 & $-$144 & 32 & $-$252 &  48 \\
  55321.364 & 0.523 & $-$124 & 60 & $-$184 &  62 \\
  55312.319 & 0.648 & $-$97 &  84 & $-$239 & 102 \\
  55320.179 & 0.622 & $-$148 & 69 & $-$147 &  80 \\
  55322.295 & 0.230 & $-$109 & 48 & $-$111 &  64 \\
  55323.311 & 0.003 & $-$39  & 34 &  $-$43 &  46 \\
  55324.285 & 0.744 & $-$107 & 43 & $-$124 &  64 \\
  55325.307 & 0.520 & $-$189 & 45 & $-$228 &  59 \\
  55326.341 & 0.306 & $-$142 & 53 & $-$176 &  70 \\
  55327.340 & 0.066 &  $-$19 & 36 &  $-$15 &  48 \\
  55328.356 & 0.838 &  $-$62 & 74 &  $-$46 &  99 \\
  55334.314 & 0.366 & $-$143 & 32 & $-$132 &  74 \\
  55341.279 & 0.661 & $-$217 & 58 & $-$235 &  74 \\
  55348.125 & 0.864 & $-$120 & 96 & $-$306 & 130 \\
  55354.296 & 0.555 & $-$201 & 55 & $-$229 &  82 \\
%\tableline
\hline\\[-7pt]
\multicolumn{6}{c}{HD\,176386}\\
%\tableline
\hline\\[-7pt]
  54610.272 & 0.174 & $-$119 & 33 & $-$121 & 35 \\
  55312.353 & 0.145 &  $-$10 & 69 &  $-$42 & 88 \\
  55322.313 & 0.224 &  $-$41 & 65 &  $-$44 & 89 \\
  55323.247 & 0.263 &  $-$62 & 62 & $-$126 & 64 \\
  55324.308 & 0.444 & $-$190 & 62 & $-$179 & 65 \\
  55325.331 & 0.582 & $-$201 & 63 & $-$225 & 71 \\
  55326.365 & 0.732 & $-$115 & 54 & $-$119 & 68 \\
  55327.244 & 0.709 &   $-$6 & 64 &  $-$49 & 79 \\
  55328.375 & 0.967 &     99 & 59 &    111 & 79 \\
  55334.339 & 0.602 &  $-$65 & 49 &  $-$50 & 63 \\
  55335.264 & 0.631 & $-$102 & 43 & $-$144 & 54 \\
  55341.317 & 0.364 & $-$101 & 65 & $-$104 & 71 \\
  55349.128 & 0.052 &    115 & 50 &    121 & 68 \\
  55354.226 & 0.724 & $-$130 & 41 & $-$162 & 62 \\
  55355.134 & 0.733 &  $-$25 & 38 &  $-$30 & 61 \\
%\tableline
\hline
\end{tabular}
\end{center}
\end{table}

\begin{figure}
\centering
\includegraphics[width=0.48\textwidth]{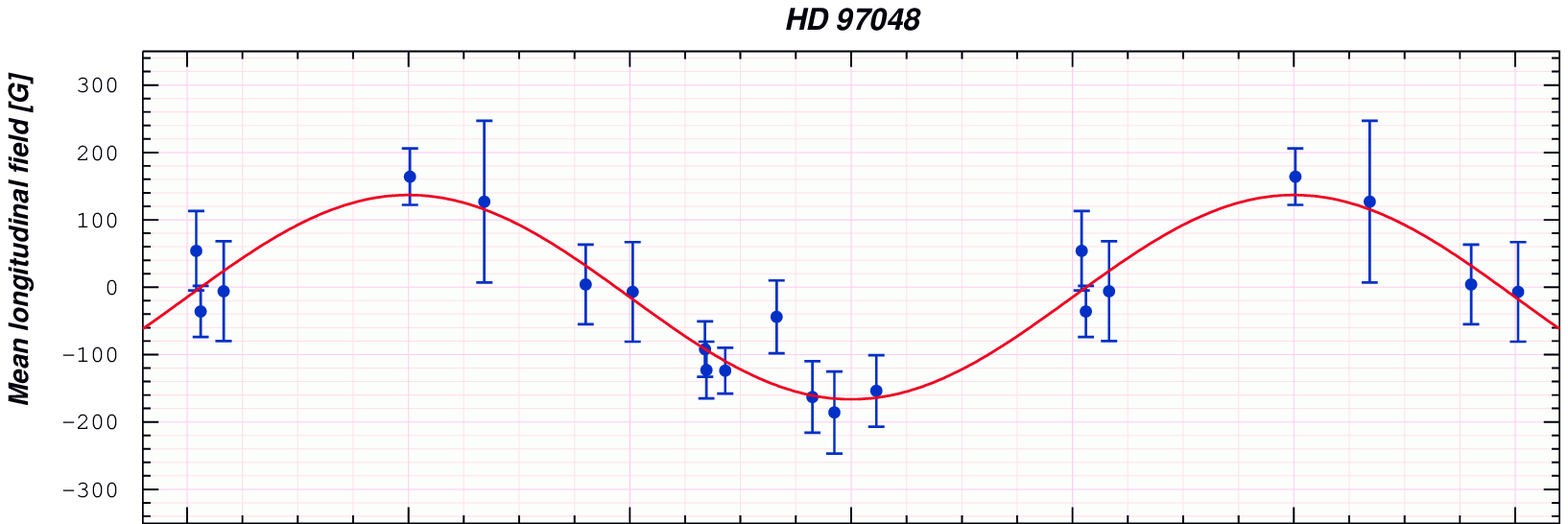}
\vskip -0.5mm
\includegraphics[width=0.48\textwidth]{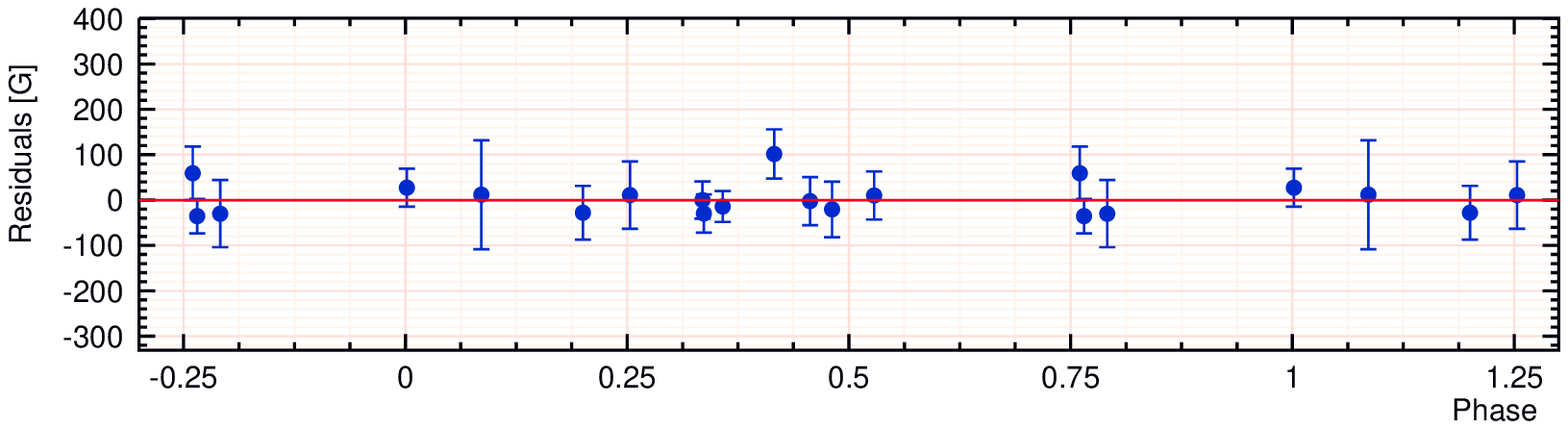}
\includegraphics[width=0.48\textwidth]{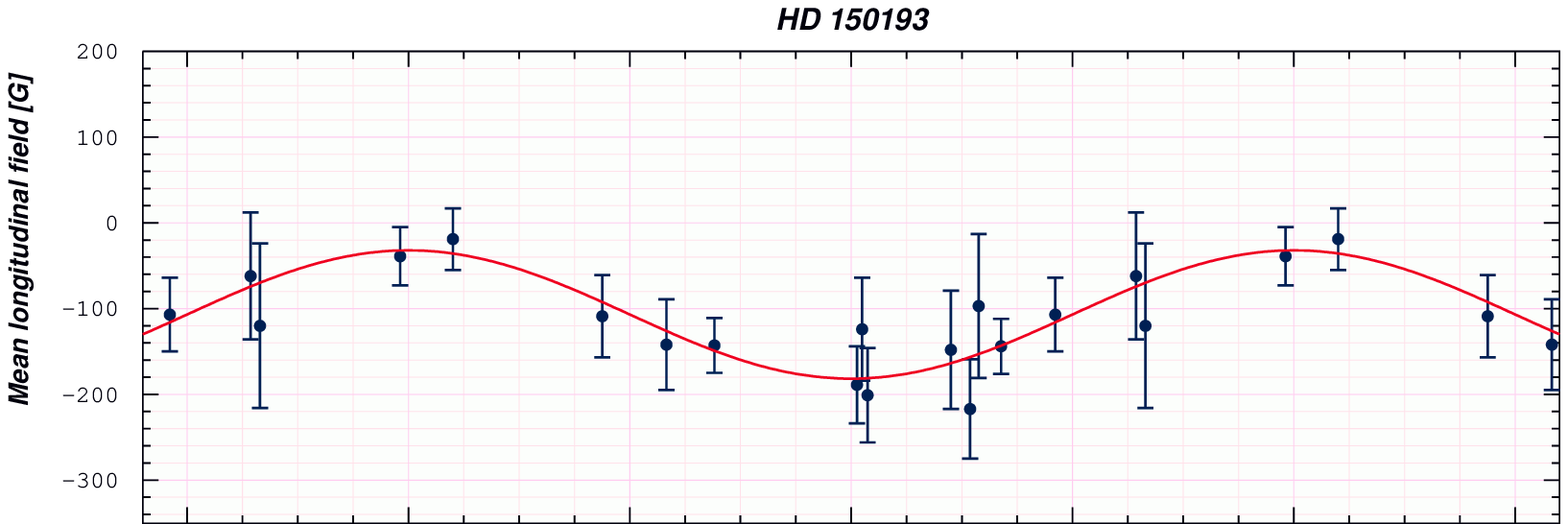}
\vskip -0.5mm
\includegraphics[width=0.48\textwidth]{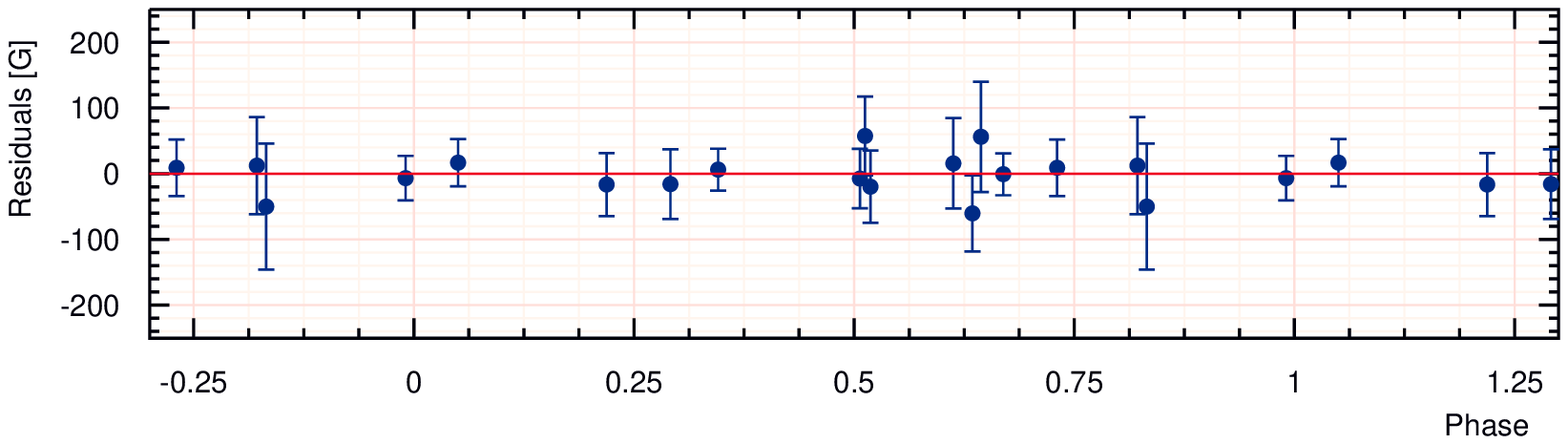}
\includegraphics[width=0.48\textwidth]{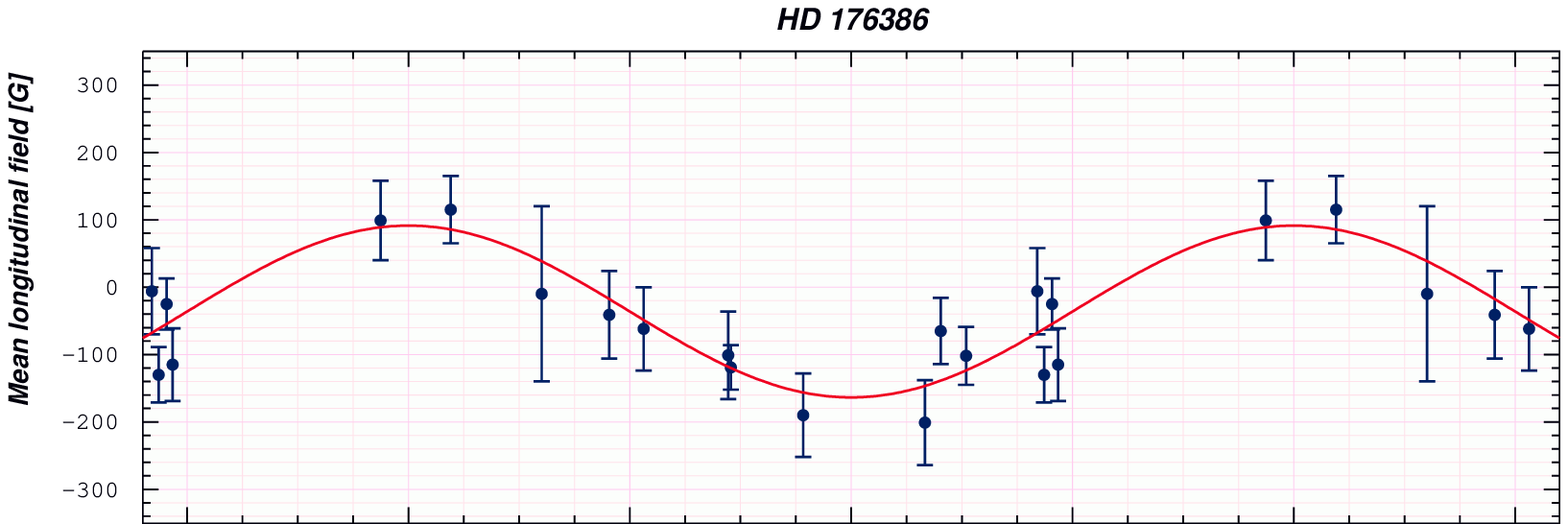}
\vskip -0.5mm
\includegraphics[width=0.48\textwidth]{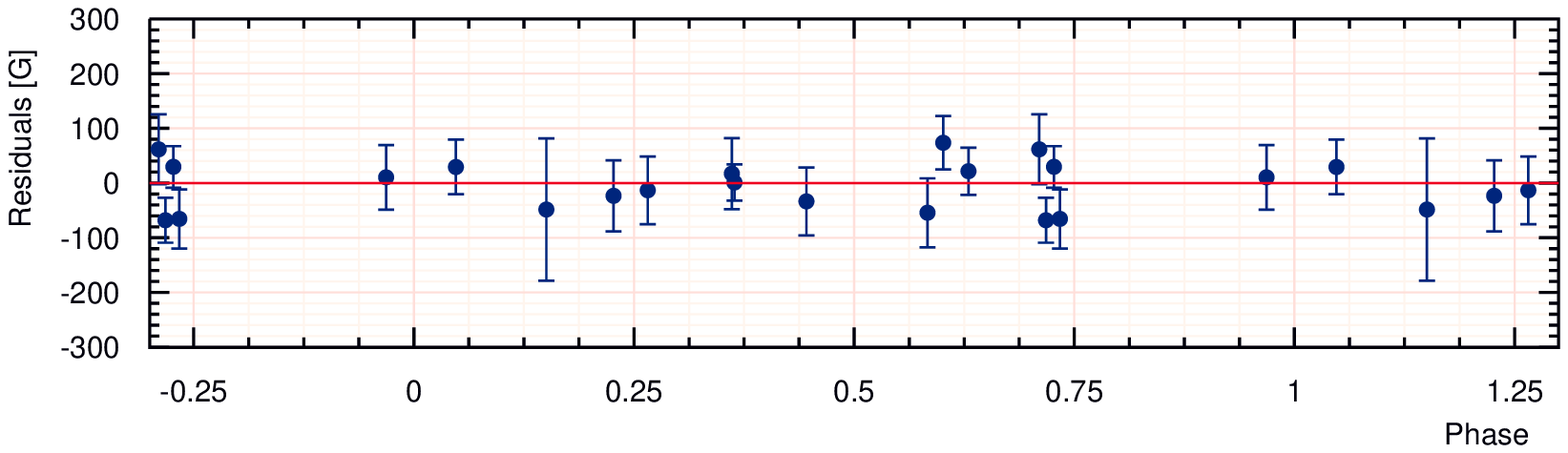}
\caption{
Phase diagrams with the best sinusoidal fit for the longitudinal magnetic field measurements of
HD\,97048 (top),
HD\,150193 (middle),
and HD\,176386 (bottom).
The residuals (observed -- calculated) are shown in the lower panel. The deviations are mostly of the same 
order as the error bars, and no systematic trends are obvious, which justifies a single sinusoid as a 
fit function. 
%rms=42.02  rchi2=0.708.
}
\label{fig:1}
\end{figure}

The targets for which we present new spectropolarimetric measurements are
listed in Table~\ref{tab:targetlist}.
Multi-epoch polarimetric spectra of the Herbig Ae/Be stars HD\,97048, HD\,150193, and 
HD\,176386 were obtained with FORS\,2\footnote{
The spectropolarimetric capabilities of FORS\,1 were moved to
FORS\,2 in 2009.
}
on Antu (UT1) from 2010 March 23 to 2010 June 7 in service mode.
Using a slit width of 0\farcs4, the spectral resolving power of FORS\,2 achieved
with the GRISM 600B is about 2000.
 A detailed description of the assessment of the longitudinal 
magnetic-field measurements using FORS\,2 was presented in our previous papers 
(e.g., Hubrig et al.\ \cite{Hubrig2004a, Hubrig2004b}, and references therein). 
%We repeat here the major steps of the magnetic field determination. 
The mean longitudinal 
magnetic field, $\left< B_{\rm z}\right>$, was derived using 

\begin{equation} 
\frac{V}{I} = -\frac{g_{\rm eff} e \lambda^2}{4\pi{}m_ec^2}\ \frac{1}{I}\ 
\frac{{\rm d}I}{{\rm d}\lambda} \left<B_{\rm z}\right>, 
\label{eqn:one} 
\end{equation} 

\noindent 
where $V$ is the Stokes parameter that measures the circular polarisation, $I$ 
is the intensity in the unpolarised spectrum, $g_{\rm eff}$ is the effective 
Land\'e factor, $e$ is the electron charge, $\lambda$ is the wavelength, $m_e$ the 
electron mass, $c$ the speed of light, ${{\rm d}I/{\rm d}\lambda}$ is the 
derivative of Stokes~$I$, and $\left<B_{\rm z}\right>$ is the mean longitudinal magnetic 
field. 

Three spectropolarimetric observations of MWC\,480 with ${\rm S/N} \ge200$ were obtained on 2009 December 30 and 2010 
December 14, and 2010 December 25 
%with S/N$>$250
with the low-resolution camera ($R=\lambda/\Delta\lambda\approx30000$)
of the echelle spectrograph SOFIN (Tuominen et al.\ \cite{Tuominen1999})
mounted at the Cassegrain focus of the Nordic Optical Telescope (NOT).
We used a 2K Loral CCD detector to register 40 echelle orders partially covering
the range from 3500 to 10\,000\,\AA{} 
with a length of the spectral orders of about 140\,\AA{} at 5500\,\AA{}.
The polarimeter is located in front of the entrance slit of the spectrograph and consists of a fixed 
calcite beam splitter aligned along the slit and a rotating super-achromatic quarter-wave plate. Two spectra 
polarised in opposite senses are recorded simultaneously for each echelle order providing sufficient separation 
by the cross-dispersion prism.
Usually two sub-exposures with the quarter-wave plate angles separated 
by $90^\circ$ are necessary to derive circularly polarised spectra.
The spectra are reduced with the 4A software package (Ilyin \cite{Ilyin2000}). 
Bias subtraction, 
master flat-field correction, 
scattered light subtraction, and weighted extraction of spectral orders comprise the standard steps of the image 
processing. A ThAr spectral lamp is used for wavelength calibration, taken before and after each target exposure 
to minimise temporal variations in the spectrograph. 
%The diagnostic potential of high-resolution circularly polarised spectra using the moment technique has been 
%discussed at length in numerous papers by Mathys (e.g.\ \cite{Mathys1993}). Wavelength shifts between right and left 
%circularly polarised spectra can in the weak-line approximation be interpreted in terms of a longitudinal 
%magnetic field  $\left<B_z\right>$.
Wavelength shifts between right- and left-hand side circularly polarised spectra were
interpreted in terms of a longitudinal magnetic field $\left<B_{\rm z}\right>$,
%%using the formalism in the framework of moment technique described by Mathys (\cite{Mathys1994}).
using the moment technique described by Mathys (\cite{Mathys1994}).

%The main limitation on the accuracy achieved in our determinations is set by the small
%number of lines that can be used for magnetic field measurements.
%The diagnosis of the 
%quadratic field is more difficult than that of the longitudinal magnetic field, and it 
%depends much more critically on the
%%size of the set
%number of lines that can be employed.

%\begin{figure}
%\centering
%\includegraphics[width=0.45\textwidth]{HD97048.pc.ps}
%\vskip -0.5mm
%\includegraphics[width=0.45\textwidth]{HD97048.pr.ps}
%\caption{
%Phase diagram with the best sinusoidal fit for the longitudinal magnetic field measurements of HD\,97048.
%The residuals (Observed -- Calculated) are shown in the lower panel. The deviations are mostly of the same 
%order as the error bars, and no systematic trends are obvious, which justifies a single sinusoid as a 
%fit function. 
%%rms=42.02  rchi2=0.708.
%}
%\label{fig:1}
%\end{figure}

%\begin{figure}
%\centering
%\includegraphics[width=0.45\textwidth]{HD150193.pc.ps}
%\vskip -0.5mm
%\includegraphics[width=0.45\textwidth]{HD150193.pr.ps}
%\caption{
%Phase diagram with the best sinusoidal fit for the longitudinal magnetic field measurements of HD\,150193.
%%rms=26.197  rchi2=0.318571.
%}
%\label{fig:2}
%\end{figure}

%\begin{figure}
%\centering
%\includegraphics[width=0.45\textwidth]{HD176386.pc.ps}
%\vskip -0.5mm
%\includegraphics[width=0.45\textwidth]{HD176386.pr.ps}
%\caption{
%Phase diagram with the best sinusoidal fit for the longitudinal magnetic field measurements of HD\,176386.
%%rms=48.9846  rchi2=0.918549.
%}
%\label{fig:3}
%\end{figure}

The frequency analysis  was performed on the longitudinal magnetic field 
measurements 
%presented in Table~\ref{tab:log_meas} including previous FORS\,1 measurements 
using a non-linear least squares fit to the multiple harmonics utilizing the Levenberg-Marquardt 
method (Press et al.\ \cite{Press92}) with an optional possibility of pre-whitening the trial harmonics.  
%All measurements including previous FORS\,1 measurements are presented in Table~\ref{tab:log_meas}.
To detect the most probable period, we calculated the frequency spectrum and for each trial frequency we performed a 
statistical F-test of the null hypothesis for the absence of periodicity (Seber \cite{Seber77}).
%To detect the most probable period we calculated the frequency spectrum for the same harmonic with a number of trial 
%frequencies by solving the linear least-squares problem. 
%At each trial frequency we performed a statistical 
%test of the null hypothesis for the absence of periodicity (Seber \cite{Seber77}), i.e.\ testing that all harmonic 
%amplitudes are at zero.
The resulting F-statistics can be thought of as the total sum including covariances of the 
ratio of harmonic amplitudes to their standard deviations, i.e.\ a signal-to-noise ratio. 
%(Ilyin \cite{Ilyin2000}). 
The derived ephemeris for the detected periods are 

\noindent
\begin{eqnarray}
{\rm HD\,97048}: \left<B_{\rm z}\right>^{\rm pos~ext} &=& \nonumber \\
{\rm MJD}55318.41923 &\pm& 0.01372 + 0.69334 \pm 0.00039 E  \nonumber \\
%\nonumber \\
{\rm  HD\,150193}: \left<B_{\rm z}\right>^{\rm pos~ext} &=& \nonumber \\
{\rm MJD}55318.05538 &\pm& 0.03072 + 1.31697 \pm 0.00013 E  \nonumber \\
%\nonumber \\
{\rm HD\,176386}: \left<B_{\rm z}\right>^{\rm pos~ext} &=& \nonumber \\
{\rm MJD}55325.70578 &\pm& 0.02185 + 0.89920 \pm 0.00008 E  \nonumber 
\end{eqnarray}

The logbook of the new FORS\,2 and the old FORS\,1 
spectropolarimetric observations is presented
in Table~\ref{tab:log_meas}.
In the first column, we indicate the MJD value at mid exposure. The phases of the measurements of 
the magnetic field
are listed in Column~2. In Columns~3 and 4, we present the longitudinal magnetic 
field $\left<B_{\rm z}\right>_{\rm all}$ measured using the whole spectrum and the 
longitudinal magnetic field 
$\left<B_{\rm z}\right>_{\rm hyd}$ using only the hydrogen lines. 
All quoted errors are 1$\sigma$ uncertainties.
The FORS\,1 measurements  
published by Hubrig et al.\ (\cite{Hubrig2009}) are presented in the first line, respectively.
Phase diagrams of the data folded with the determined periods are presented in Fig.~\ref{fig:1}.
%Figs.~\ref{fig:1} -~\ref{fig:3}.
The quality of our fits is described by a reduced $\chi^2$-value 
%which appears in the four panels of Fig.~\ref{fig:all}.
that is 0.71 for HD\,98048, 0.32 for HD\,150193, and  
0.92 for HD\,176386.

%the spectral range from 3290\,\AA{} to 4500\,\AA{} in the 
%blue arm and the spectral range  from  5680\,\AA{} to 9460\,\AA{} 
%in the red arm. The slit width was set to $0\farcs{}3$ for the red arm
%and $0\farcs{}4$ for the blue arm, corresponding to a resolving power 
%of $\lambda{}/\Delta{}\lambda{} \approx$ 110,000 and 
%$\approx$ 90,000, respectively.

\begin{table*}
\centering
\caption{
Magnetic field models constrained by the FORS\,1/2 spectropolarimetric observations.
}
\label{tab:dipvals}
\begin{tabular}{cc|r@{$\pm$}lr@{$\pm$}lr@{$\pm$}l}
\hline
\hline
\multicolumn{2}{c|}{Object} &
\multicolumn{2}{c}{HD\,97048} & 
\multicolumn{2}{c}{HD\,150193} & 
\multicolumn{2}{c}{HD\,176386} \\
%\multicolumn{2}{c}{33\,Eri} \\
\hline
$\overline{\left< B_{\rm z}\right>}$ {\rule{0pt}{2.6ex}} & [G]           & $-$14.8 & 14.3 & $-$106.9&  7.4 & $-$36.2 &  15.9 \\
$A_{\left< B_{\rm z}\right>}$        & [G]           & 151.6 & 19.6 & 74.8& 9.8 & 127.5 & 24.7 \\
$v$\,sin\,$i$                        & [km s$^{-1}$] & \multicolumn{2}{c}{} & \multicolumn{2}{c}{} & 165& 8 \\
$R$                                  & [R$_{\odot}$] & \multicolumn{2}{c}{} & \multicolumn{2}{c}{} & 3.0 & 0.5 \\
$v_{\rm eq}$                         & [km s$^{-1}$] & \multicolumn{2}{c}{} & \multicolumn{2}{c}{} & 169 & 28 \\
$i$                                  & [$^{\circ}$]  & 42.8 & 2.5 & 38 & 9 & 78 & 46 \\
$\beta$                              & [$^{\circ}$]  & 84.8 & 5.0 & 41.8 & 10.1 & 37 & 106 \\
%B$_{\rm d}$                         & [G]           & 5300 & 1100 & 570 &  190 & 3800 &  700 & 1500 &  400 \\
B$_{\rm d}$                          & [G]           & 720 & 120 & 590 & 40 & 700 & 650 \\
\hline
\end{tabular}
\end{table*}

All three Herbig Ae/Be stars exhibit a single-wave variation in 
the longitudinal magnetic field during the stellar rotation cycle.
These observations are usually considered as evidence 
of a dominant dipolar contribution to the magnetic field topology. 
%Our previous study of the abundances of HD\,101412 using UVES and HARPS spectra
% found that $v \sin i  = 3\pm 1$\,km\,s$^{-1}$ (Cowley et al.\ \cite{Cowley2010}).
Assuming that the studied Herbig Ae/Be stars are  oblique dipole rotators,
the magnetic dipole axis tilt $\beta$ is constrained by 
$\left< B_{\rm z}\right>^{\rm max}/\left< B_{\rm z}\right>^{\rm min} = \cos(i+\beta)/\cos(i-\beta)$,
where the inclination angle $i$ is derived from resolved observations of their disks.
In Table~\ref{tab:dipvals}, we show for each star in rows~1 and 2 
the mean value $\overline{\left< B_{\rm z}\right>}$ and the amplitude of the field variation $A_{\left< B_{\rm z}\right>}$.
For the star HD\,97048, Lagage et al.\ (\cite{Lagage2006}) used VISIR observations of the
mid-infrared emission of polycyclic hydrocarbons at the surface of the disk to determine
$i=42.8\pm2.5^{\circ}$.
Using this value, we obtain a magnetic obliquity of $\beta=84.8\pm5.0^{\circ}$ and the dipole strength 
$B_{\rm d}=720\pm120$\,G.
The disk inclination of HD\,150193, $i=38\pm9^{\circ}$, was determined by Fukagawa et al.\ (\cite{Fukagawa2003}) using 
Subaru near-infrared imaging.
With this value, we find a magnetic obliquity $\beta=42\pm10^{\circ}$ and 
the dipole strength $B_{\rm d}=590\pm40$\,G.

\begin{figure}
\centering
\includegraphics[angle=270,totalheight=0.30\textwidth]{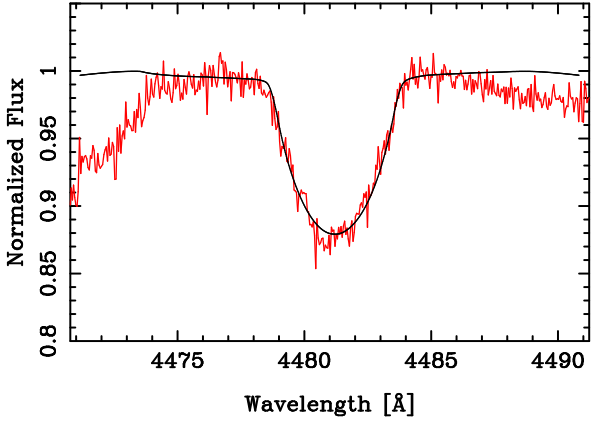}
\caption{
Determination of the $v$\,sin\,$i$ value for HD\,176386.
The line profile of \ion{Mg}{ii}~4481 in the NTT/EMMI spectrum is fitted by the corresponding
synthetic spectrum assuming $T_{\rm eff}=9900$\,K and log\,$g=3.7$.
}
\label{fig:4}
\end{figure}

\begin{figure}
\centering
\includegraphics[angle=270,totalheight=0.30\textwidth]{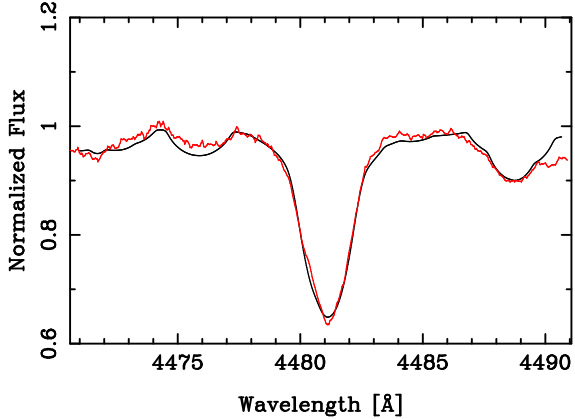}
\caption{
Determination of the $v$\,sin\,$i$ value for MWC\,480.
The line profile of \ion{Mg}{ii}~4481 in the SOFIN spectrum is fitted by the corresponding
synthetic spectrum assuming $T_{\rm eff}=8500$\,K and log\,$g=3.5$.
}
\label{fig:chucka}
\end{figure}

For HD\,176386, the inclination angle $i$ is not known because both IR (Siebenmorgen et al.\ \cite{Siebenmorgen2000}) and far-UV 
spectroscopy (Martin-Za\"idi et al.\ \cite{MartinZaidi2008}) suggest that the circumstellar 
matter has the form of an envelope rather than the form of a disk.
To constrain the magnetic field geometry of this star, we determined that
$v$\,sin\,$i=165\pm8$\,km\,s$^{-1}$ from our NTT/EMMI spectrum obtained for this star in 1995 July 8. In Fig.~\ref{fig:4},
we present the line profile
of \ion{Mg}{ii}~4481 overplotted by the corresponding synthetic spectrum assuming
$T_{\rm eff}=9900$\,K and log\,$g=3.7$, and 
solar abundances. We note that the literature values for $T_{\rm eff}$ widely differ from each other, from $T_{\rm eff}=10715$\,K 
as given by van den Ancker et al.\ (\cite{Ancker1997}) to $T_{\rm eff}=9527$\,K derived by Meyer \& Wilking (\cite{Meyer2006}).
The equatorial rotation velocity can be calculated using the relation
$v_{\rm eq}=50.6$\,$R/P$, where $R$
%%=11.2$\pm$1.4\,R$_\odot$
is the stellar radius in solar units and P is the rotation period in days. 
%%Using this formula, we obtain $v_{\rm eq}$ = 41$\pm$5\,km\,s$^{-1}$. 
%The radius of HD\,176386 $R$=3$\pm$0.5\,R$_\odot$ was estimated using the
%Stefan--Boltzmann law and information on bolometric luminosity and effective temperature from van den Ancker et al.\ \cite{Ancker1997}
%and Meyer \& Wilking {\cite{Meyer2006}).
%For these stars we estimated the radii using the
%Stefan--Boltzmann law. The information on bolometric luminosity and effective temperature of the three Herbig stars 
%was found in the recent paper of van der Plas et al.\ (\cite{vanderPlas2008}), while
Taking into account the uncertainties in atmospheric parameter 
determinations for HD\,176386, we adopt the radius $R=3\pm0.5$\,$R_\odot$.
With this radius and the period $P=0.899$\,d, we
%a pre-main sequence star with $T_{\rm eff}$=9900\,K and log\,$g$ =3.7
%of the age 3--4 Myr 
obtain $v_{\rm eq}=169\pm28$\,km\,s$^{-1}$, which leads to the poorly constrained
$i=78\pm46^{\circ}$,
magnetic obliquity $\beta=37\pm106^{\circ}$, and the dipole strength $B_{\rm d}=700\pm650$\,G.
The parameters of the magnetic field dipole models for all three stars are listed in the last two rows of Table~\ref{tab:dipvals}.

%The inclination angles of disks of Herbig Ae/Be stars (which are expected to be identical
%to the inclination angle
%of the stellar rotation axis) can be reliably derived only for resolved observations of disks.
%Fedele et al.\ (\cite{Fedele2008}) used VLTI/MIDI observations to determine $i$=80$\pm$7$^{\circ}$.
%Using this value, we obtain $v_{\rm eq}$=3$\pm$1\,km\,s$^{-1}$. 
%Assuming that HD\,101412 is an oblique dipole 
%rotator,  we follow the definition of Preston (\cite{Preston1967})
%\begin{equation} 
%r = \frac{\left< B_{\rm z}\right>^{\rm min}}{\left< B_{\rm z}\right>^{\rm max}} 
%  = \frac{\cos \beta \cos i - \sin \beta \sin i}{\cos \beta \cos i + \sin \beta 
%\sin i}, 
%\end{equation} 
 %\noindent 
%so that the obliquity angle $\beta$ is given by
 %\begin{equation} 
%\beta =  \arctan \left[ \left( \frac{1-r}{1+r} \right) \cot i \right]. 
%\label{eqn:4} 
%\end{equation} 
%From the phase curve determined above, with $\left< B_{\rm z} \right>^{\rm max}$=474$\pm$32\,G
%and $\left< B_{\rm z} \right>^{\rm min}$=$-$456$\pm$32\,G, we find $r$=$-$0.962$\pm$0.075, 
%which for $i$=80$\pm$7$^{\circ}$ leads to a magnetic obliquity of $\beta$=84$\pm$13$^{\circ}$. 

\begin{table}
\centering
\caption{
Magnetic field measurements of MWC\,480 with SOFIN.
%In the first line we list the earlier FORS\,1 measurement 
%published by Hubrig et al.\ (\cite{Hubrig2006}).
%Phases are calculated according to the ephemeris of 
%${\rm JD} = 2455402.46 + 13.893~{\rm E}$.
%All quoted errors are 1$\sigma$ uncertainties.
}
\label{tab:log_sofin}
\begin{tabular}{cr@{$\pm$}l}
\hline
\multicolumn{1}{c}{MJD} &
%\multicolumn{1}{c}{Phase} &
\multicolumn{2}{c}{$\left<B_{\rm z}\right>$ [G]} \\
%\multicolumn{2}{c}{$\left<B_{\rm z}\right>_{\rm hyd}$} \\
%& \multicolumn{2}{c}{[G]} \\
%\multicolumn{2}{c}{[G]} \\
\hline
\hline
53296.350 & 87  & 22 \\
%\hline
55195.086 & $-$952 & 177 \\
55544.090 &  458   & 149 \\
55555.982 &  189   & 204 \\
\hline
\end{tabular}
\end{table}

The logbook of our new SOFIN spectropolarimetric observations of MWC\,480 is presented in Table~\ref{tab:log_sofin}.
In the first column, we list the MJD
values for the middle of the spectropolarimetric observations and in Col.~2 we present the mean longitudinal magnetic 
field $\left<B_{\rm z}\right>$.
In the first line, we list the earlier FORS\,1 measurement 
published by Hubrig et al.\ (\cite{Hubrig2006}).
All quoted errors are 1$\sigma$ uncertainties.
Since only four magnetic field measurements are available, no search for a periodicity
can be undertaken.
On the other hand, the rotation period can be estimated using the known inclination angle 
$i=36\pm1^{\circ}$ (Pi\'etu et al.\ \cite{Pietu2006}), the radius $R=1.67$\,$R_\odot$ (Blondel \& Tijn A Djie \cite{Blondel2006}),
and $v$\,sin\,$i=90\pm2$\,km\,s$^{-1}$ estimated on our SOFIN spectra.
In Fig.~\ref{fig:chucka}, we present the line profile
of \ion{Mg}{ii}~4481 overplotted by the synthetic spectrum 
assuming $T_{\rm eff}=8500$\,K and log\,$g=3.5$.
With this value for $v$\,sin\,$i$, we obtain
$v_{\rm eq}=153\pm8$\,km\,s$^{-1}$. 
By substituting these values into the relation $P=50.6$\,$R/v_{\rm eq}$,
it follows that $P=0.531\pm0.029$\,d.
This period is probably very close to
the true rotation period, since Grady et al.\ (\cite{Grady2010})
report that their FUSE observations integrated over the 
individual FUSE orbits indicate a marginal detection of variability
on the timescale of 0.5\,d.

\section{Magnetic field and CS environment of MWC\,480}

In our earlier studies based on the low-resolution polarimetric FORS\,1 spectra, 
we reported the detection of distinct Zeeman features in several Herbig Ae/Be stars at the positions of 
the \ion{Ca}{ii} H\&K lines 
(Hubrig et al.\ \cite{Hubrig2004c,Hubrig2006,Hubrig2007,Hubrig2009}).
%These features
The \ion{Ca}{ii} H\&K lines
frequently display variable multi-component complex structures in both the Stokes~$V$
and Stokes~$I$ spectra and it has been suggested that they originate
 at the base of the stellar wind, as well as in 
the gaseous accretion flow. 
The star MWC\,480 exhibits a significant spectral variability, which implies that there is a 
strong interaction between the star and its CS environment.
The temporal behaviour of the complex variable structure of H$\alpha$, \ion{He}{i}, and the \ion{Na}{i}~D line
was monitored in several previous  studies (e.g., Beskrovnaya \& Pogodin \cite{Beskr2004}; Kozlova et al.\ \cite{Kozlova2007}),
which showed that the stellar wind is non-homogeneous and consists of several components, 
differing from each other in terms of their velocities.
Furthermore, structural variations in the stellar wind have been 
found to correlate with changes in the accretion process in the envelope.  

\begin{figure}
\centering
\includegraphics[width=0.40\textwidth]{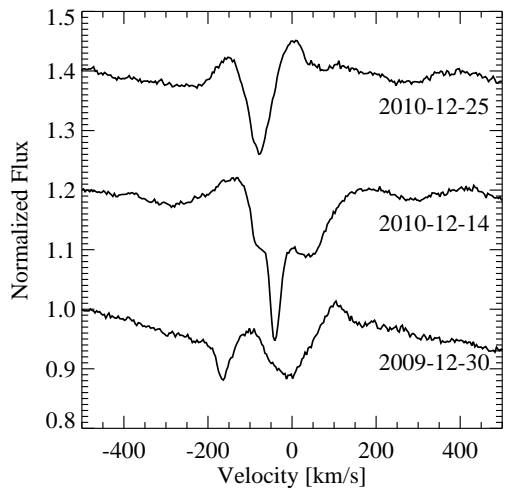}
\caption{
The behaviour of the \ion{Fe}{ii} line at $\lambda$5018.4 at the three observational epochs
2009 December 30 (bottom),
2010 December 14 (middle),
and 2010 December 25 (top).
}
\label{fig:fe_velocity}
\end{figure}

In our high-resolution SOFIN spectra, the contamination of the spectra by the CS material is clearly visible.
In Fig.~\ref{fig:fe_velocity}, we show the presence of a sharp blue component in the Mult.~42 \ion{Fe}{ii} line
at $\lambda$5018.4 during the first and the second observational epochs in 2009 and 2010.
During the last observation, the wings of the \ion{Fe}{ii} line appear in emission.
%with a sharp violet component in all three phases. 
In addition, two other lines, $\lambda\lambda$5169.0 and 4923.9 
of the same multiplet, exhibit a similar behaviour. The shift of this sharp blue component,
%component, which certainly belongs to the CS environment, 
reaches $\sim$153\,km\,s$^{-1}$ during the first SOFIN observation in 2009,
%reaching $\sim$153\,km\,s$^{-1}$. 
and is of the same order as the shifts measured for the blue 
components of the \ion{Na}{i}~D,
the \ion{Ca}{ii} H\&K lines, and the cores of the hydrogen lines. 
%In addition, the width of this blue component 
%is comparable to the width of the blue components of the \ion{Na}{i} doublet.

%\begin{figure}
%\centering
%\includegraphics[width=0.45\textwidth]{hd31648h.eps}
%\caption{
%The behaviour of the \ion{Fe}{ii} line at $\lambda$5018.4 on three observational epochs.
%}
%\label{fig:Fevelocity}
%\end{figure}

%\begin{figure}
%\centering
%\includegraphics[width=0.33\textwidth]{hd101IHa.eps}
%\includegraphics[width=0.33\textwidth]{hd101IHb.eps}
%\caption{
%UVES and HARPS spectra of HD\,101412 obtained at different epochs in spectral regions 
%around the H$\alpha$ line (top) and the H$\beta$ line (bottom). The spectra are presented with MJD dates increasing 
%from bottom to top and offset in vertical direction for clarity.
%Please note that H$\beta$ falls into the gap of the UVES\,DIC2~390+760 setting.
%}
%\label{fig:3}
%\end{figure}

\begin{figure*}
\centering
\includegraphics[width=0.30\textwidth]{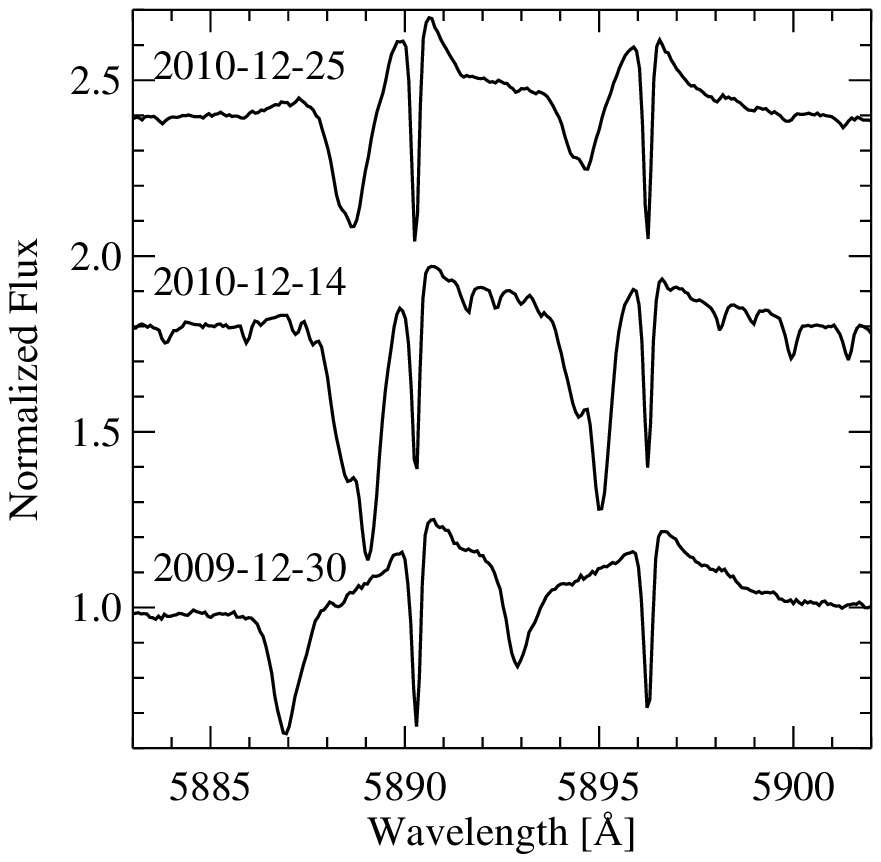}
\includegraphics[width=0.30\textwidth]{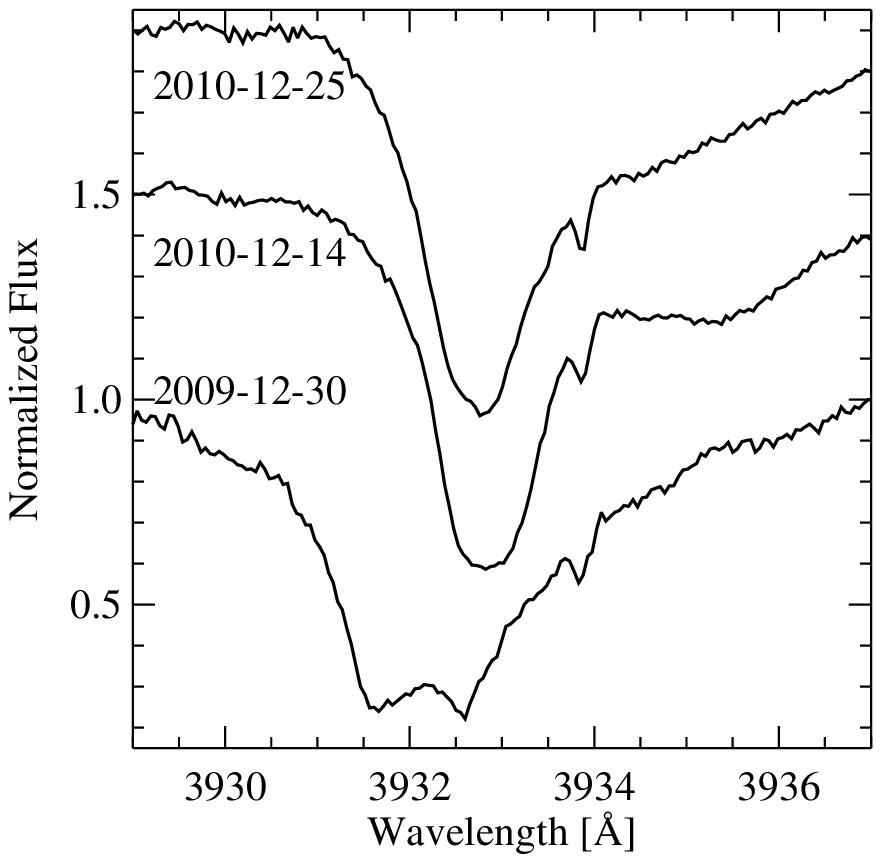}
\includegraphics[width=0.30\textwidth]{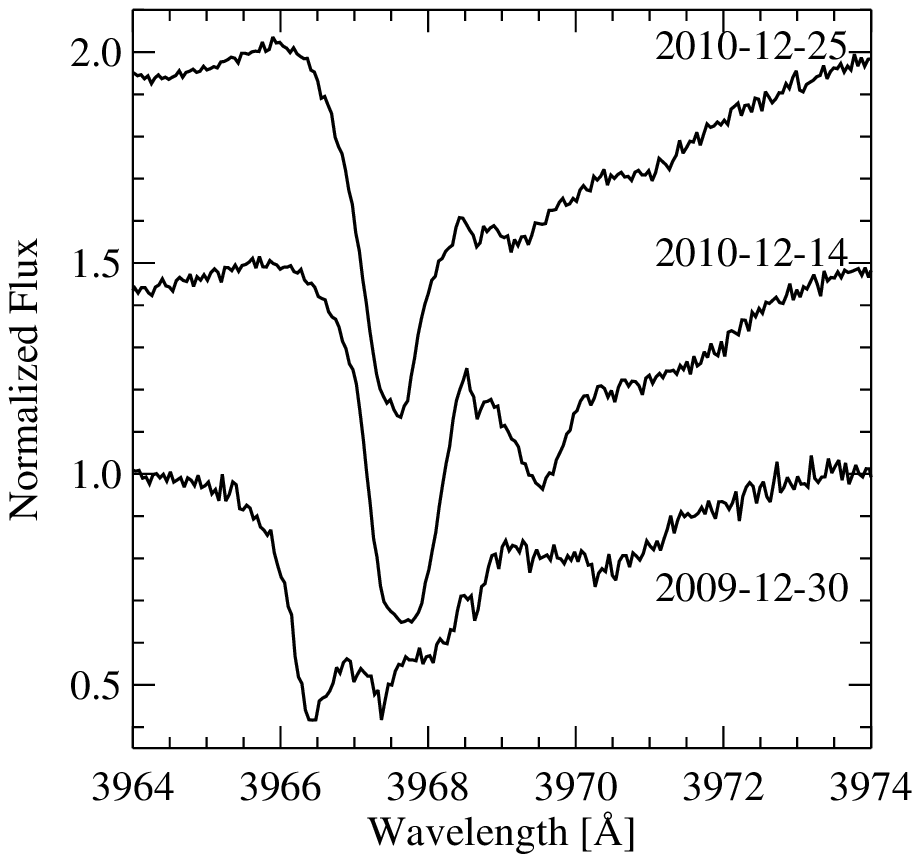}
\caption{
The  variability and complex structure of \ion{Na}{i}~D (left) and \ion{Ca}{ii} K\&H
(middle and right) in the spectra
%The  variability and complex structure of the \ion{Na}{i} D lines in the Stokes~$I$ and Stokes~$V$ spectra
of MWC\,480 at the same observational epochs as in Fig.~\ref{fig:fe_velocity}.
}
\label{fig:na_i}
\end{figure*}

%\begin{figure}
%\centering
%\includegraphics[width=0.40\textwidth]{herrbigav.eps}
%\caption{
%The  variability and complex structure of \ion{Na}{i} doublet on three observational epochs.
%}
%\label{fig:na_v}
%\end{figure}

%\begin{figure}
%\centering
%\includegraphics[width=0.40\textwidth]{herrbigIVa.eps}
%\caption{
%The presence of variable Zeeman features at the position of the CS \ion{Na}{i} doublet lines.
%}
%\label{fig:na_iv}
%\end{figure}

%\begin{figure}
%\centering
%%\includegraphics[width=0.40\textwidth]{herrbigb.eps}
%%\includegraphics[width=0.40\textwidth]{herrbigbv.eps}
%\includegraphics[width=0.35\textwidth]{herrbig3IVb.eps}
%\caption{
%The variability and complex structure of the \ion{Ca}{ii} K line in the Stokes~$I$ and Stokes~$V$ spectra
%of MWC\,480 at the same observational epochs as in Fig.~\ref{fig:fe_velocity}.
%}
%\label{fig:ca_i}
%\end{figure}

%\begin{figure}
%\centering
%%\includegraphics[width=0.40\textwidth]{herrbigc.eps}
%%\includegraphics[width=0.40\textwidth]{herrbigcv.eps}
%\includegraphics[width=0.35\textwidth]{herrbig3IVc.eps}
%\caption{
%The variability and complex structure of the \ion{Ca}{ii} H line in the Stokes~$I$ and Stokes~$V$ spectra
%of MWC\,480 at the same observational epochs as in Fig.~\ref{fig:fe_velocity}.
%}
%\label{fig:ca_v}
%\end{figure}

%\begin{figure*}
%\centering
%\includegraphics[width=0.40\textwidth]{herrbigIVb.eps}
%\includegraphics[width=0.40\textwidth]{herrbigIVc.eps}
%\caption{
%The presence of variable Zeeman features at the position of the CS \ion{Ca}{ii} doublet lines. 
%}
%\label{fig:ca_iv}
%\end{figure*}

In agreement with previous studies, the \ion{Na}{i}~D lines are highly variable, but 
the \ion{Ca}{ii} H\&K lines also show a complex structure consisting of several components and pronounced variability. 
The behaviour of the \ion{Na}{i} and \ion{Ca}{ii} doublets
at three different observation epochs is presented in Fig.~\ref{fig:na_i}.
Owing to strong blending, it is impossible to measure the shift between the line profiles in right- and left-hand side
circularly polarised spectra for all line components. The compilation of 
measurements of the component positions in the \ion{Ca}{ii} and \ion{Na}{i} doublet lines and our attempts to determine
the longitudinal magnetic field are presented in Table~\ref{tab:fields}.
%in Appendix A. 
%Our attempts to determine the shifts in terms of the longitudinal field are presented in Table~\ref{~\ref{tab:fields}. 
%The blue CS components in the \ion{Na}{i} D
%lines appear well separated at the first epoch allowing us to measure a weak longitudinal magnetic field of the order of $\sim$$-$100\,G. 

Weak magnetic fields of the order of several tens of Gauss were detected in the CS components of 
the \ion{Na}{i} and \ion{Ca}{ii} doublet lines and
can probably be considered as CS magnetic
fields. If we assume a dipole configuration for the magnetic field geometry,
then we expect to measure a magnetic field strength of $-$76\,G detected in the CS component of the \ion{Ca}{ii} K line 
at a distance of $\sim$2.3 stellar radii.
This consideration is fully in line with the generally assumed scenario of magnetospheric
accretion in Herbig Ae and T\,Tauri stars. The stellar magnetic field truncates the accretion disk at a few stellar radii and 
gas accretes along magnetic channels from the protoplanetary disk to the star. The disk is truncated at a radius R$_{\rm T}$, which 
depends upon the strength of the surface magnetic field, the mass accretion radius, the stellar mass,
 and the stellar radius 
(e.g.\ Bouvier et al.\ \cite{Bouvier2007}).
%In the framework of the scenario of magnetospheric accretion
%the disk is truncated at a radius R$_{\rm T}$, which depends upon the strength of the surface magnetic field, the mass accretion radius, 
%the stellar amss and the stellar radius (Bouvier et al.\ \cite{Bouvier2007}). At this radius and interior to it the idsk material 
%is expected to be  
For a dipole configuration of the magnetic field geometry, a magnetic field strength of $-$120\,G detected in the CS component 
of the \ion{Na}{i}~D is expected to be measured at a distance of $\sim$2.0  stellar radii. 

\section{Discussion}

Using FORS\,2 low-resolution multi-epoch polarimetric spectra, we have determined
rotation/magnetic periods for the three Herbig Ae/Be stars HD\,97048, HD\,150193, and HD\,176386.
In the framework of the rigid rotator model usually assumed for magnetic stars
on the upper main-sequence,
the period of the magnetic field variation corresponds to the stellar rotation period. Our study indicates that dipole 
models provide a satisfactory fit to the acquired magnetic data. 
%Still, additional future spectropolarimetric 
%observations are needed to confirm the detected periodicity. 

\begin{figure}
\centering
\includegraphics[width=0.45\textwidth]{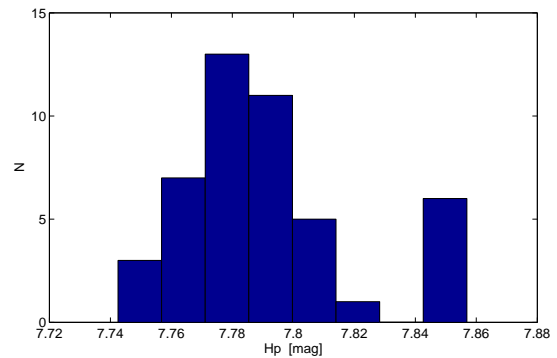}
\caption{
Available {\em Hipparcos} measurements for MWC\,480 with outliers observed during HJD 8674.75-8674.94. 
} 
\label{fig:hist}
\end{figure}

\begin{figure}
\centering
\includegraphics[width=0.45\textwidth]{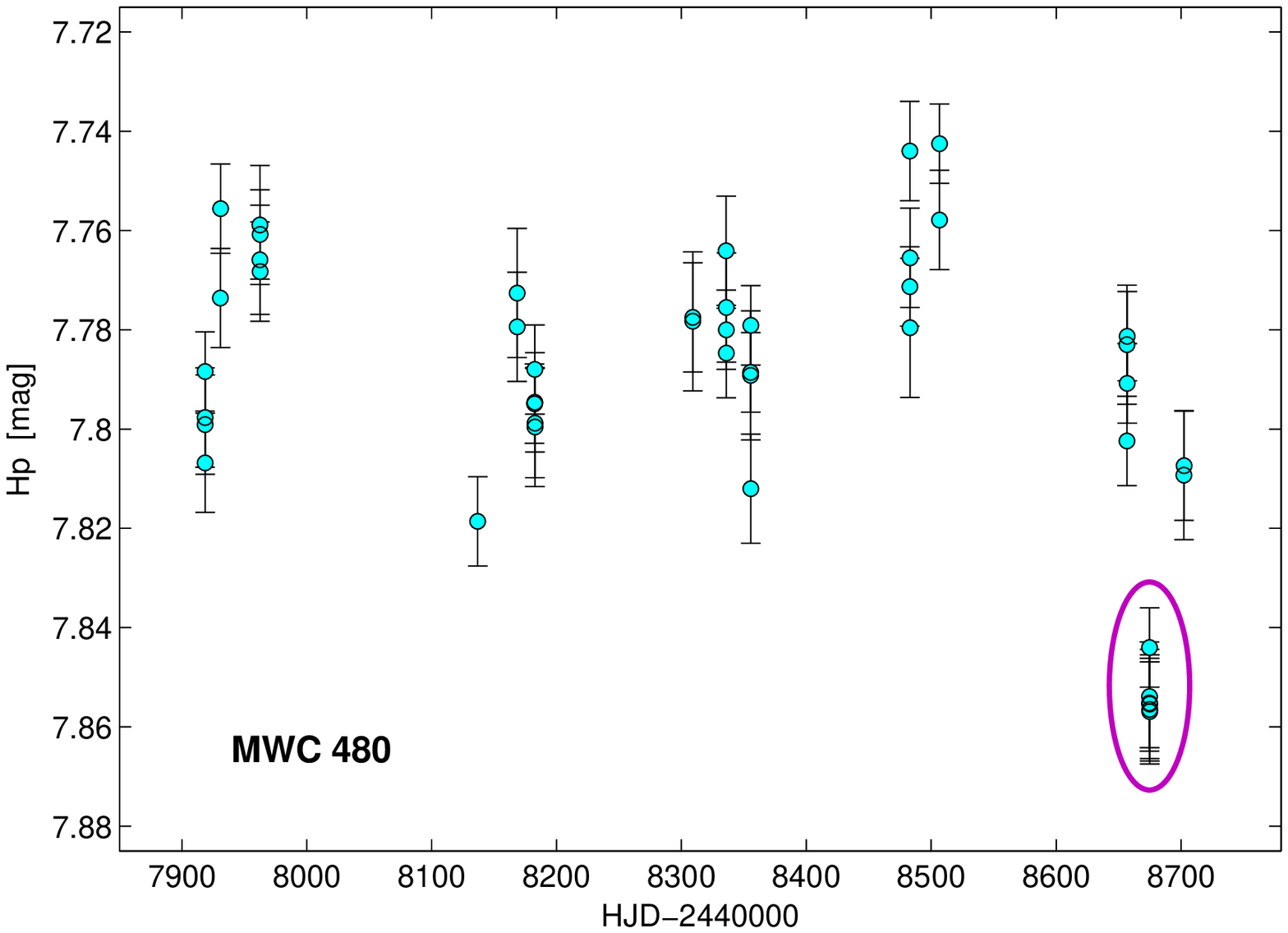}
\includegraphics[width=0.45\textwidth]{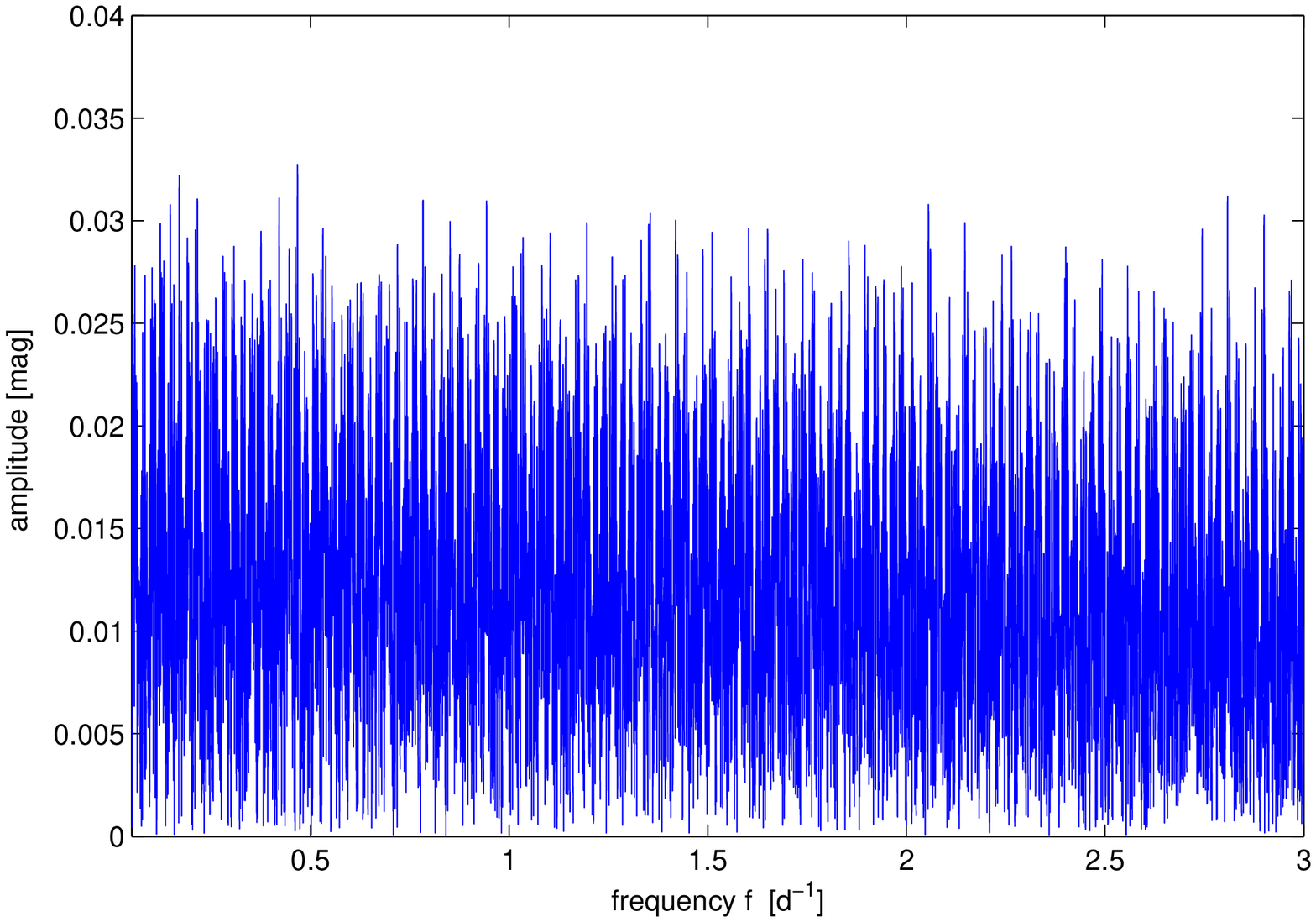}
\caption{
{\em Hipparcos} observations of MWC\,480 and the corresponding frequency periodogram.
The six outliers are indicated by an ellipse.
}
\label{fig:zdena}
\end{figure}

On the other hand, our search for rotation-modulated photometric
variability using the available photometric databases has been unsuccessful: no clear variation was found for the three 
stars in either the ASAS database (Pojma\'nski \cite{pojm02}) or {\em Hipparcos} photometry. 
%As we mention above, the FUSE data suggest a presence of the periodicity of 0.5\,d in the star MWC\,480 (Grady et al.\ \cite{Grady2010}). 
The star MWC\,480 was not observed by ASAS,
and only {\em Hipparcos} photometry is available for a period search. It displayed moderate light 
changes of up to 0.098\,mag on timescales from tens of hours to several days. The behaviour of MWC\,480 resembles the characteristics 
typical of other photometrically monitored Herbig Ae/Be stars, for which light variations are likely of stochastic nature and 
caused by the 
pre-main-sequence disk accretion phenomena (e.g.\ Rucinski et al.\ \cite{Rucinski2010}). 
Using archive {\em Hipparcos} data, we have found an indication of a variability with a period of 1.6\,d.
However, it seems that six outliers appear
during the period HJD~8674.75--8674.94.
This group of outliers is located on the right side of the histogram presented in Fig.~\ref{fig:hist}. 
Excluding these outliers from the analysis,
we found no indication for periodicity.
In Fig.~\ref{fig:zdena}, we present {\em Hipparcos}
observations of MWC\,480 and the corresponding frequency periodogram. 
%$\Omega$ quantifies the significance of the particular period:
%$\Omega = \Sigma( m_p(t_i)-m_{\rm mean})^2 w_i \ \Sigma w_i$ 
%%- hence [mag^2]$, 
%where $m_p(t_i)$ is the
%predicted value of the magnitude for an individual time $t_i$, $m_{\rm mean}$, is the mean
%detrended magnitude for the set, and $w_i$ is the weight of the individual
%measurement. 

\begin{figure}
\centering
\includegraphics[width=0.45\textwidth]{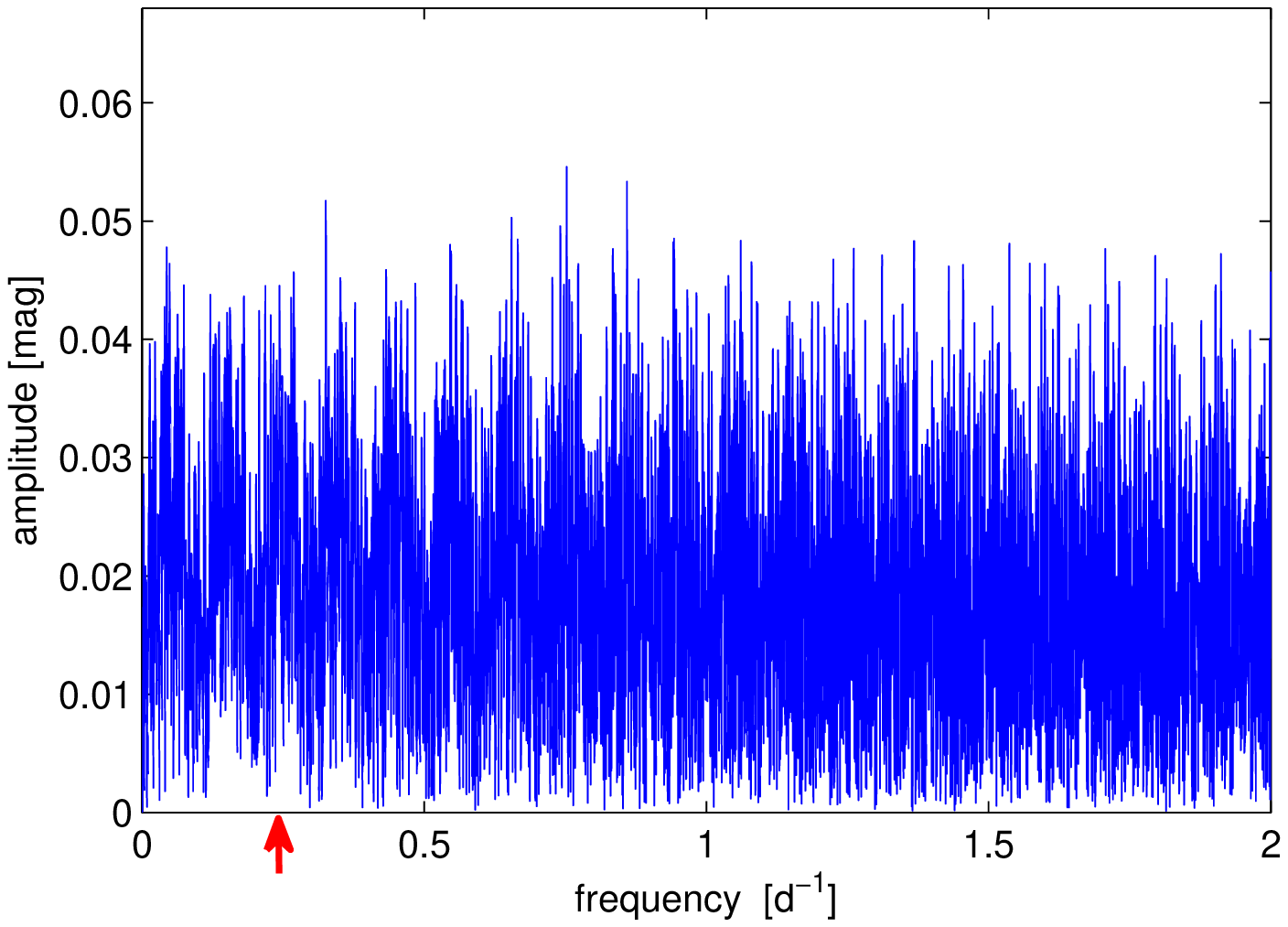}
\includegraphics[width=0.45\textwidth]{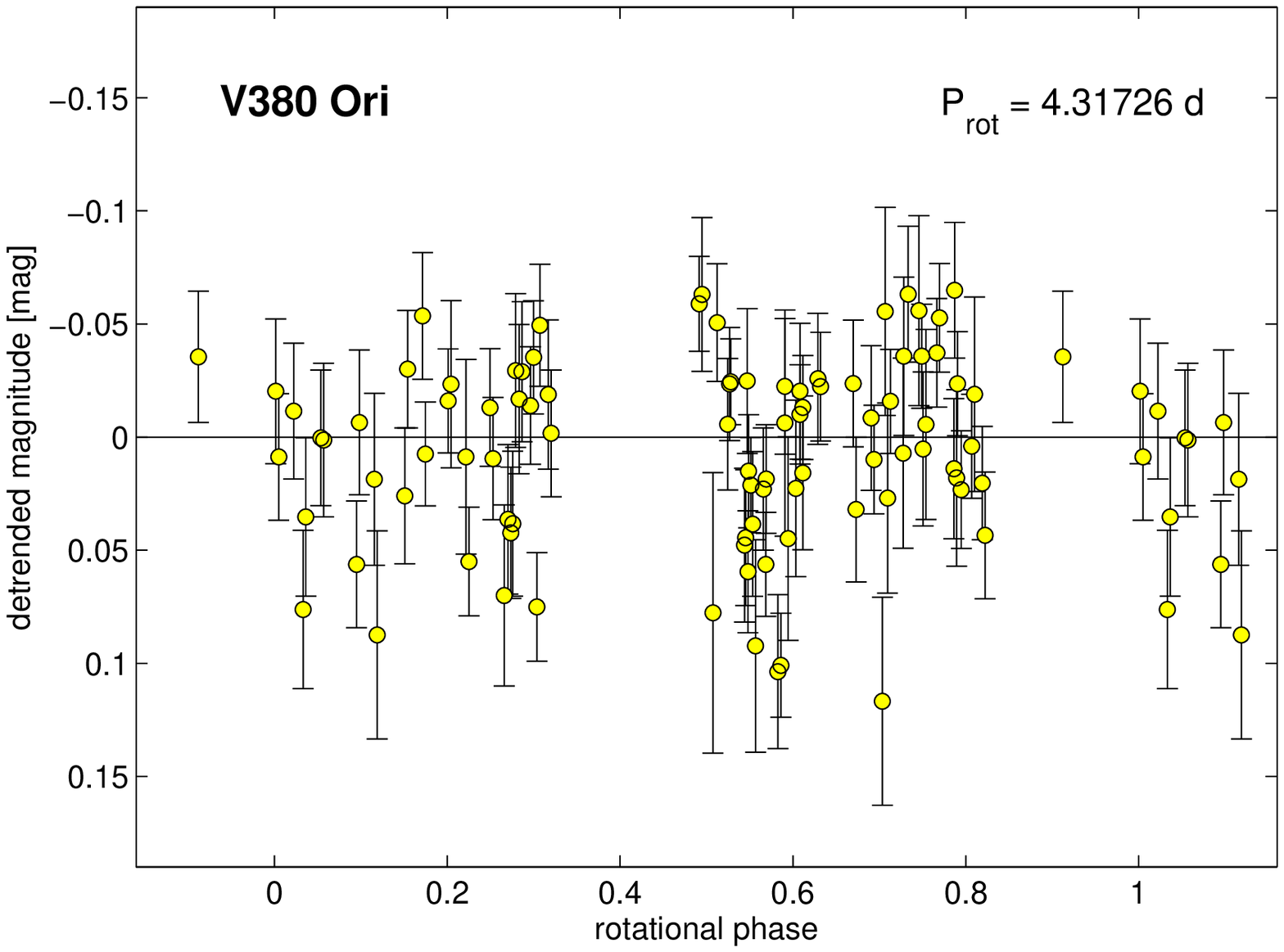}
\caption{
%{\em Hipparcos} observations of V380\,Ori and the corresponding frequency periodogram. 
Frequency periodogram from {\em Hipparcos} observations of V380\,Ori.
%The red arrow in this figure denotes the rotation period  of V380\,Ori determined by Alecian et al.\ (\cite{Alecian2009}). 
The red arrow in this figure denotes the rotation period determined by Alecian et al.\ (\cite{Alecian2009}). 
The lower panel presents the corresponding phase diagram. No periodicity is detected.
}
\label{fig:zdenb}
\end{figure}

Kozlova et al.\ (\cite{Kozlova2007}) studied the spectral 
variability of MWC\,480 in the region of H$\alpha$. 
They found an equivalent width variability on 
a timescale of about 1200\,d and suggest that the most plausible reason for the detected cyclic variability is the 
reconstruction of the inner structure of the CS gas envelope, caused by the presence of a low mass 
companion or planet around the star.
It is not yet clear how and whether this periodicity is related to the appearance of the red-shifted component
in the \ion{Ca}{ii} H\&K lines during the second and third epochs of the SOFIN observations.

Our experience shows that the photometric periodic variability of Herbig Ae stars is generally difficult to detect.
As an example, we show in Fig.~\ref{fig:zdenb} {\em Hipparcos} photometric data for V380\,Ori,
for which the rotation period $P=4.31276\pm0.00042$\,d
was determined from magnetic field measurements (Alecian et al.\ \cite{Alecian2009}). As we show in this figure, no periodicity 
can be detected in the photometric data.

%We note that in the past
Until recently, rotation periods had been known for
only three Herbig Ae stars, V380 Ori, HD\,101412, and HD\,135544B.
They had been determined from direct measurements of longitudinal
magnetic fields (for V380\,Ori and HD\,101412), photometry (for HD\,101412), or radial velocities (for HD\,135544B)
(Alecian et al.\ \cite{Alecian2009}; Hubrig et al.\ \cite{Hubrig2011}; M\"uller et al.\ \cite{Muller2011}).

%The red arrow in this figure denotes the rotation period determined by Alecian et al.\  
The longitudinal magnetic fields of HD\,97048, HD\,150193, and HD\,176386 are not very strong compared to the kG fields detected in 
V380\,Ori, HD\,101412, and Z\,CMa (Alecian et al.\ \cite{Alecian2009}; Hubrig et al.\ \cite{Hubrig2010,Hubrig2011}; 
Szeifert et al.\ \cite{Szeifert2010}).
%Regarding the strength of the magnetic field of MWC\,480, 
Grady et al.\ (\cite{Grady2010}) placed limits on the field strength for MWC\,480 assuming that this star
is accreting magnetospherically.
%{\bf Grady et al.\ (\cite{Grady2010}) estimated the average magnetic 
%field strength for MWC\,480 by equating the field energy to that of accretion plasma flow.}
%required for the coupling between the accreting plasma and the stellar magnetic field in the scenario of magnetospheric
%accretion.
%% to lift the accreting material from the disk midplane to higher stellar latitudes near the stellar photosphere. 
Their result, $B_{\rm aver}=3^{+5}_{-2}$\,kG, appears comparable 
with our measurements presented in this work.

The detection of a weak longitudinal magnetic field in the blue-shifted CS components of the \ion{Na}{i} and 
the \ion{Ca}{ii} doublet lines in the polarimetric spectra of MWC\,480 implies that  for Herbig Ae/Be stars
the Doppler-shifted spectropolarimetric contributions
from photospheric and circumstellar environmental diagnostic lines should be investigated using high-resolution
polarimetric observations over the rotation period. These studies will allow us to apply 
the technique of Doppler Zeeman tomography to determine the correspondence between the magnetic field structure and 
the radial density and temperature profiles.
%Multi-epoch observations at a resolution of 100\,000 over the rotation period will allow us to determine the 
%magnetic field topology in more detail than it was ever done before. 
%The Doppler-shifted spectral contributions
%from photospheric and circumstellar environmental diagnostic lines will enable us to apply the 
%technique of Doppler Zeeman tomography to determine the correspondence between the magnetic field structure and the radial dens
%ity
%and temperature profiles. For the first time we will attempt to reconstruct two-dimensional velocity and magnetic field maps
%from a set of phase-resolved emission and absorption line profiles.

%high resolution HARPS spectra, we will be able
%for the first time to model the Stokes~$V$ profiles of the observed Zeeman signatures not only in photospheric
% lines but also 
%in various wind and accretion diagnostic lines. 
%Furthermore, knowledge of the magnetic field structure combined with the 
%determination of chemical composition and 
%surface element distribution are indispensable to constrain theories
%on star formation and magnetospheric accretion.
{
\acknowledgements
This research has made use of the SIMBAD database,
operated at CDS, Strasbourg, France.
MAP and RVY acknowledge the support obtained by RFBR grant No\,07-02-00535a and Sci.Schole No\,6110.2008.2.
}

\appendix

\section{Magnetic field measurements of components identified in the \ion{Ca}{ii} and \ion{Na}{i} doublet line profiles.}

During the first epoch of our SOFIN observations
in December 2009, the Ca doublet lines were found to have three components.
Two of them are shifted to the blue by 146.4\,km\,s$^{-1}$ and 83.1\,km\,s$^{-1}$,
respectively, and the third one is shifted to the red by 13.7\,km\,s$^{-1}$. This velocity corresponds to the 
radial velocity of the photospheric spectral lines.
During the epochs 2 and 3, an additional strong broad component appears on the red side.
The shift measured during the second epoch is almost 100\,km\,s$^{-1}$.
Such a strong broad red-shifted component is also discovered in the higher number Balmer
lines, providing potential evidence of a lower mass companion.
In any case, future studies should focus on 
multi-epoch spectroscopic observations to monitor the radial velocities and the line profile variations in this star and prove the presence of such a companion.
Wavelength shifts for MWC\,480 and the corresponding velocities and magnetic field measurements of
its  components
identified in the \ion{Ca}{ii} and \ion{Na}{i} doublet line profiles are presented in Table~\ref{tab:fields}.
%Variable distinct Zeeman features are well visible at the positions of a few components,
%but due to strong blending it is not possible to measure the shift between the line profiles in right- and left-hand side
%circularly polarised spectra for all components. 
%%Our attempts to determine the shifts in terms of the longitudinal field are presented in Table~\ref{~\ref{tab:fields}. 
The blue CS components in the \ion{Na}{i}~D
lines appear to be clearly separated at the first epoch allowing us to measure a weak longitudinal magnetic field of the order of $\sim$$-$100\,G. 
The \ion{Na}{i}~D lines themselves are in emission.
They have overlapping sharp absorption components.
%The position of these sharp absorption components does not change from one epoch to the next. 
During the second and third epochs of observations,
the blue CS components indicate a double structure and are located closer to the emission \ion{Na}{i}~D lines. 
%The compilation of 
%measurements of the component positions in the \ion{Ca}{ii} and \ion{Na}{i} doublet lines and our attempts to determine the shifts 
%between the line profiles in right- and left-hand side
%circularly polarised spectra in terms of the longitudinal magnetic field are presented 
%in Table~\ref{tab:fields} in Appendix A. 
\begin{table*}
\caption[]{
Wavelength shifts for MWC\,480 with corresponding velocities and magnetic field measurements of components
identified in the \ion{Ca}{ii} and \ion{Na}{i} doublet line profiles.
Numbers with colons indicate the most uncertain values due to blends.
}
\begin{center}
\begin{tabular}{c|rrr|rrr|rrr}
\hline
\hline
\multicolumn{1}{c|}{} &
\multicolumn{3}{c|}{Epoch 1} &
\multicolumn{3}{c|}{Epoch 2} &
\multicolumn{3}{c}{Epoch 3} \\
\multicolumn{1}{c|}{Component} &
\multicolumn{1}{c}{$\Delta\lambda$} &
\multicolumn{1}{c}{velocity shift} &
\multicolumn{1}{c|}{$\left<B_{\rm z}\right>$} &
\multicolumn{1}{c}{$\Delta\lambda$} &
\multicolumn{1}{c}{velocity shift} &
\multicolumn{1}{c|}{$\left<B_{\rm z}\right>$} &
\multicolumn{1}{c}{$\Delta\lambda$} &
\multicolumn{1}{c}{velocity shift} &
\multicolumn{1}{c}{$\left<B_{\rm z}\right>$} \\
\multicolumn{1}{c|}{} &
\multicolumn{1}{c}{[\AA{}]} &
\multicolumn{1}{c}{[km\,s$^{-1}$]} &
\multicolumn{1}{c|}{[G]} &
\multicolumn{1}{c}{[\AA{}]} &
\multicolumn{1}{c}{[km\,s$^{-1}$]} &
\multicolumn{1}{c|}{[G]} &
\multicolumn{1}{c}{[\AA{}]} &
\multicolumn{1}{c}{[km\,s$^{-1}$]} &
\multicolumn{1}{c}{[G]} \\
\hline
\multicolumn{10}{c}{\ion{Ca}{ii} 3933.7} \\
\hline
1$^{\rm st}$ & $-$1.92 &$-$146.4 &       &$-$0.86 &$-$65.6 &17 &  & & \\
2$^{\rm nd}$ & $-$1.09 & $-$83.1 &  $-$24   & & & &$-$0.84 & $-$64.1 & \\
3$^{\rm rd}$ &    0.18 &  13.7 &  $-$30     &0.18 & 13.7  & & 0.18 & 13.7& \\
4$^{\rm th}$ &         &       &       & 1.31: & 99.9: & & & & \\
\hline
\multicolumn{10}{c}{\ion{Ca}{ii} 3968.5} \\
\hline
1$^{\rm st}$ & $-$1.92 & $-$146.4 &$-$76 &$-$0.85 &$-$64.3 & 56 & & & \\
2$^{\rm nd}$ & $-$1.09 & $-$83.1 &$-$20  & & & & & & \\
3$^{\rm rd}$ &           &       & & 0.17 &13.0 & &0.17 &13.0 & \\
4$^{\rm th}$ &         &       & & 1.06:& 80.1: & & 0.67: & 50.7:& \\
\hline
\multicolumn{10}{c}{H$\epsilon$} \\
\hline
            & 0.19 & 14.4 & $-$706 & & & & & &\\
\hline
\multicolumn{10}{c}{\ion{Na}{i} 5890.0}  \\
\hline
1$^{\rm st}$ & $-$3.01 & $-$152.8 &$-$116 &$-$1.43 & $-$72.8 &  & &  \\
2$^{\rm nd}$ &         &       &       &$-$0.90 & $-$45.8 & & $-$1.31 &$-$66.7 &  \\
3$^{\rm rd}$  &    0.28 &  14.3  &$-$16  &0.28    &14.3    & $-$21 & 0.28    &14.3 &  \\
\hline
\multicolumn{10}{c}{\ion{Na}{i} 5895.9}  \\
\hline
1$^{\rm st}$ & $-$2.99 &$-$152.1 &  $-$90 &$-$1.44 & $-$72.8 & & &   \\
2$^{\rm nd}$ &         &       &        & $-$0.90& $-$45.8&  &$-$1.3 &$-$66.2 &  \\
3$^{\rm rd}$  &    0.30 &  15.2  &$-$120  & 0.30 &15.2 & & 0.30 & 15.2 & \\
\hline
\end{tabular}
\end{center}
\label{tab:fields}
\end{table*}

\begin{figure}
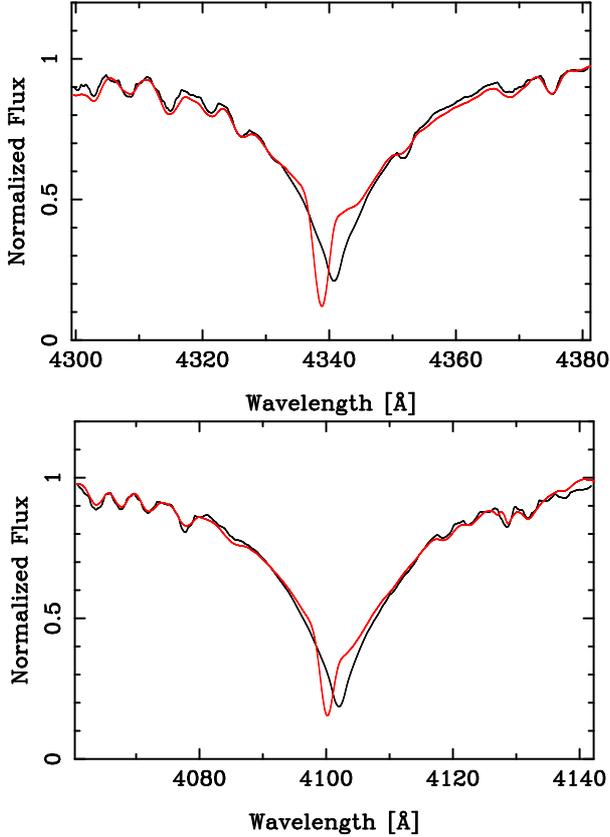

\centering
\includegraphics[angle=270,totalheight=0.30\textwidth]{Gam85g35.ps}
\includegraphics[angle=270,totalheight=0.30\textwidth]{Del85g35.ps}
\caption{
The H$\gamma$  (upper plot) and H$\delta$ (lower plot) line profiles in the spectrum of MWC\,480 observed with FORS\,1 in 2004 
with an overplotted
synthetic spectrum assuming  $T_{\rm eff}=8500$\,K and log\,$g=3.5$.
}
\label{fig:chuckb}
\end{figure}

As for the large shifts in the blue components of the \ion{Na}{i}~D lines and 
the \ion{Ca}{ii} H\&K lines detected in the SOFIN spectrum in 2009, the cores of the hydrogen lines
from H$\beta$ to H$8$ in our FORS\,1 spectra obtained in 2004 are also shifted 
to the blue by $\sim$150\,km\,s$^{-1}$.
In Fig.~\ref{fig:chuckb}, we show the behaviour of the H$\gamma$ and  H$\delta$ line profiles
together with the calculated synthetic spectra for atmosphere model parameters $T_{\rm eff}=8500$\,K and log\,$g=3.5$.
%The hydrogen cores are shifted to the blue by $\sim$150\,km\,s$^{-1}$. 
Interestingly,  this pattern is the opposite of that seen for the Herbig Ae star HD\,101412,
where the higher order Balmer lines display red shifts that decrease towards
the lower order Balmer lines.
It is possible that the opposite trend for hydrogen line-core velocities in
both stars can be explained by different viewing angles, i.e.\
we view MWC\,480 mostly in the wind direction at an angle $i=36\pm1^{\circ}$ (Pi\'etu et al.\ \cite{Pietu2006}), 
whereas HD\,101412 is viewed in the direction of the accretion flow with $i=80\pm7^{\circ}$ (Fedele et al.\ \cite{Fedele2008}).

\end{document}